\documentclass[aps,prb,twocolumn,amsmath,amssymb,gallery,footinbib,superscriptaddress,floatfix]{revtex4-2}
\usepackage{graphicx}
\usepackage{dcolumn}
\usepackage{bm}
\usepackage{hyperref}
\usepackage{stackrel,amssymb}
\usepackage{placeins}
\hypersetup{colorlinks=true, linkcolor=blue, citecolor=blue, urlcolor=magenta, pdftitle={On the search of displacive chiral phase transitions}, pdfauthor={F. Gomez-Ortiz}}
\usepackage{bbding}
\usepackage{xcolor} 
\usepackage[normalem]{ulem}

\begin{document}
\preprint{APS/123-QED}
\title{Inhomogeneous Electric Fields for Precise Control and Displacement of Polar Textures}
\author{Fernando G\'omez-Ortiz}
\email[]{fgomez@uliege.be}
\affiliation{Physique Théorique des Matériaux, QMAT, Université de Liège, B-4000 Sart-Tilman, Belgium} 
\author{Louis Bastogne}
\affiliation{Physique Théorique des Matériaux, QMAT, Université de Liège, B-4000 Sart-Tilman, Belgium}
\author{He Xu}
\affiliation{Physique Théorique des Matériaux, QMAT, Université de Liège, B-4000 Sart-Tilman, Belgium}
\author{Philippe Ghosez}
\affiliation{Physique Théorique des Matériaux, QMAT, Université de Liège, B-4000 Sart-Tilman, Belgium} 

\date{\today}
\begin{abstract}
Since the discovery of polar topological textures, achieving efficient control and manipulation of them has emerged as a significant challenge for their integration into nanoelectronic devices. In this study, we use second principles molecular dynamic simulations to demonstrate the precise and reversible control of domain arrangements stabilizing diverse polarization textures through the application of various inhomogeneous electric fields. Furthermore, we conduct an in-depth study of ferroelectric domain motion under such fields, revealing features consistent with creep dynamics and establishing an upper limit for their propagation speed. Notably, our findings show that domain walls exhibit an asymmetric inertial response, present at the onset of the dynamics but absent during their cessation. These findings provide valuable insights into the dynamic behavior of polar textures, paving the way for the development of high-speed, low-power nanoelectronic applications.
\end{abstract}
\maketitle
\section{Introduction}
\label{sec:introduction}
Topological textures of ferroelectric materials have garnered substantial attention in the last years owing to their complex phase diagram~\cite{Junquera-23}, their nanometer lengthscales~\cite{Catalan-12}, their interesting functional properties such as negative capacitance~\cite{Iniguez-19} or chirality~\cite{Louis-12,Shafer-18} and their ultra-fast dynamics~\cite{Daranciang-12,Li-21}.
More recently, ferroelectric domain walls~\cite{Streiffer-02} and skyrmion bubbles~\cite{Das-19} or, more generally speaking, ferroelectric bubble domains~\cite{Zhang-17} have also emerged as promising nanoscale candidates for information carriers and with potential to revolutionize von Neumann computational approches~\cite{Meier-24}. Due to their quasi-particle behavior~\cite{Aramberri-24} as well as their scalability~\cite{Catalan-12}, stability~\cite{Das-19,Das-21,Govinden-23}, stochastic dynamics~\cite{Aramberri-24,Prokhorenko-24,Gomez-24} and controllable density~\cite{Nahas-20,Govinden-23,Aramberri-24,Prokhorenko-24} they are appealing for nanoelectronic applications such as racetrack memories~\cite{Stuart-08} or token-based computing~\cite{Pinna-18,Jibiki-20,Meier-24}. 

However, in order to exploit the potential of these topological textures for practical applications, it is essential to develop precise methods for controlling both their instantaneous polarization state and their dynamical motion.

Lately, electric field pulses applied over mesoscopic regions~\cite{Li-19,Prosandeev-22,Prosandeev-23,Zajac-24} have been used to stimulate optical phonons in the material and stabilize different hidden phases by changing the frequency of the pulses or by recurrently applying them a concrete number of times. While promising, these methods have so far stabilized only specific phases, and deterministic control of the entire phase diagram for topological textures remains elusive. Additionally, predicting the resulting polar arrangement  given the number of pulses, their duration, and the spacing between them is challenging.
Recently, we proposed an alternative method showing the deterministic and dynamical tailoring of polar topologies using spatially modulated acoustic phonon excitations ``{\sc{apex}}''~\cite{Bastogne-24}. Leveraging the strong coupling between polarization and strain in ferroelectric materials, the method uses acoustic phonons to generate strain fields that modulate the local polarization.

In this work, we demonstrate that the conceptual framework of inducing distortions with wavevectors $q\neq 0$ can be extended to spatially modulated electric fields, enabling comparable control and manipulation of polar topologies. Furthermore, we perform an in-depth study of the dynamics of ferroelectric domain walls under such inhomogeneous fields revealing features consistent with the creep dynamics~\cite{Paruch-02,Paruch-13}, unveiling the inertial response of ferroelectric domain walls, focusing on the mechanisms governing their motion and the critical velocities they can reach, which exceed those observed in their magnetic counterparts~\cite{Jiang-17,Pham-24}.
\begin{figure*}[tbp]
     \centering
      \includegraphics[width=15cm]{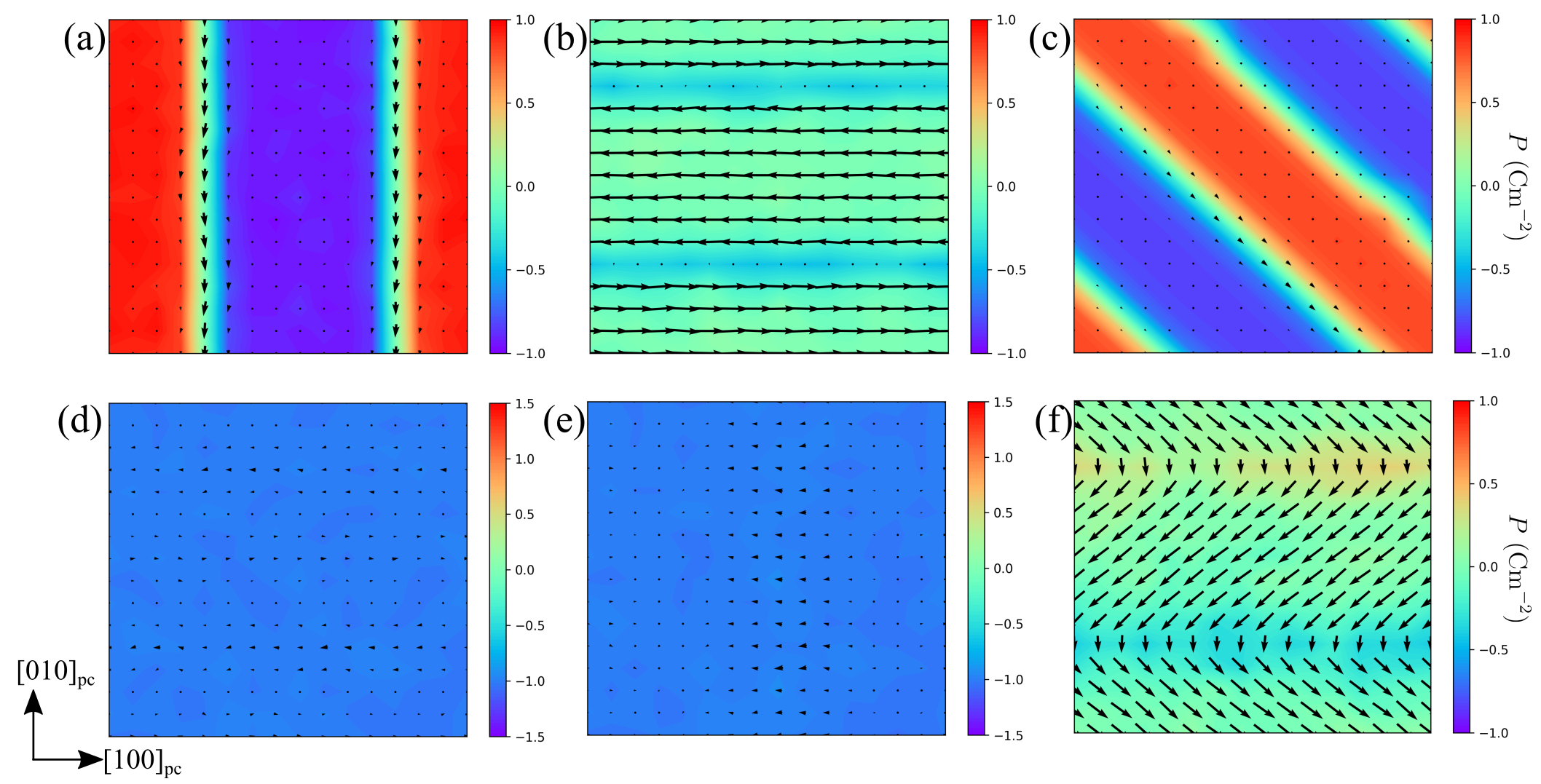}
      \caption{Stabilization of different domain textures in bulk PbTiO$_3$ after the application of an inhomogeneous electric field with the functional form of Eq.~\ref{eq:cos}. (a) $\lambda_x=16$ u.c. $\mathbf{A}=0.51\cdot u_z$ MV/cm. (b) $\lambda_y=16$ u.c. $\mathbf{A}=0.51\cdot u_x$ MV/cm. (c) $\lambda_x=\lambda_y=16$ u.c. $\mathbf{A}=0.51\cdot u_z$ MV/cm. (d) $\lambda_y=8$ u.c. $\mathbf{A}=1.54\cdot u_z$ MV/cm (e) $\lambda_x=16$ u.c. $\mathbf{A}=25.7\cdot u_x$ MV/cm (f) in-plane tensile strain $a=b=3.955$ \AA~ and $\lambda_y=16$ u.c. $\mathbf{A}=0.51\cdot u_x$ MV/cm. Arrows indicate the in-plane components of the polarization, whereas the color map represents the out-of-plane polarization.}
      \label{fig:cos} 
\end{figure*}
\section{Methods}
\label{sec:Methods}
Second-principles (SP) atomistic models were constructed using the \textsc{Multibinit}~\cite{gonze2020abinit} software by fitting data from density functional theory (DFT) calculations produced with the \textsc{Abinit}~\cite{gonze2020abinit} software package. The generalized gradient approximation (GGA) with the PBESol exchange-correlation functional and a planewave-pseudopotential approach with optimized norm-conserving pseudopotentials from the PseudoDojo server~\cite{hamann2013optimized,van2018pseudodojo} were employed, considering as valence electrons $5d^{10}6s^26p^2$ for Pb, $3s^23p^63d^24s^2$ for Ti, and $2s^22p^4$ for O. A plane-wave energy cutoff of $65\ Ha$ and an $8\times 8\times 8$ $\Gamma$-centered k-point mesh were used.
The harmonic part of the models was derived from Density Functional Perturbation Theory (DFPT) as implemented in \textsc{Abinit}~\cite{gonze2020abinit}. Dynamical matrices for the relaxed cubic $Pm\bar{3}m$ structure were computed using a $4\times 4\times 4$ q-point mesh. Various properties such as dipole-dipole interaction, Born effective charge, strain-phonon coupling, and elastic constants were extracted from the DFPT framework.
The anharmonic part of the SP potential was fitted on a training set consisting of 3644 DFT configurations. A cutoff radius of $\sqrt{3}/2$ times the cubic lattice cell parameter was used to generate the anharmonic symmetry-adapted terms (SAT) from third to eighth order, considering strain-phonon coupling and anharmonic elastic constants. With this parameter set, 29 anharmonic SATs were automatically chosen to minimize energy forces and stresses through a goal-function, and 66 additional terms were automatically produced to ensure the bounding of the SP potential.

Finite-temperature simulations were carried out using a Hybrid Molecular Dynamics-Monte Carlo approach~\cite{duane1987hybrid,betancourt2017conceptual} to study the deterministic and dynamic switching of ferroelectric domains. For the analysis of domain motion, conventional molecular dynamics (MD) simulations in the canonical ensemble were employed. A time step of $0.72$ fs was used, ensuring accurate resolution of the dynamics relative to the periods of the applied electric fields.

In order to account for the effect of the inhomogeneous electric fields, an external force is added to each atom. Since the Born effective charge tensor ${\bold Z}^*$ describes the linear relationship between the force on an atom and
the macroscopic electric field ${\bold E}$~\cite{Gonze-97}, the force induced along $\beta$ on an atom $\kappa$ at a given position ${\bf r}$ by a field in direction $\alpha$ is equal to $F_{\kappa,\beta}  = {Z}^*_{k,\alpha\beta}{E}_{\alpha}({\bf r})$.
\section{Results}
Let us first discuss the stabilization and dynamical control of different polar textures depending on the shape of the electric fields employed. We begin by focusing on simple domain structures that can be stabilized using a single modulated electric field before extending our analysis to more complex phases.

\emph{Stabilization of stripe domains.-} When the electric field has the functional form described by Eq.~(\ref{eq:cos}), where $\mathbf{A}$ dictates the amplitude of the electric field and $\bm{\lambda}$ the periodicity of the electric field in real space, stripe domains can be stabilized along different directions as shown in Fig.~\ref{fig:cos}.
\begin{equation}
    {\bold E}({\bold r})=\mathbf{A}\cdot \cos(\frac{2\pi x}{\lambda_x}+\frac{2\pi y}{\lambda_y}+\frac{2\pi z}{\lambda_z}).
    \label{eq:cos}
\end{equation}
The value of $\bm{\lambda}$ will naturally determine the direction of the modulation of the domains and their periodicity in real space whereas the value of $\mathbf{A}$ will determine the direction of the electric field. Therefore, if $\bm{\lambda}$ and $\mathbf{A}$ are parallel (or antiparallel) the electric field will promote instable head to head and tail to tail domain walls whereas if $\bm{\lambda}$ and $\mathbf{A}$ are perpendicular, the electric field will couple with transversal optical modes and stabilize regular 180$^\circ$ domain walls in tetragonal ferroelectrics.
Indeed, as shown in Fig.~\ref{fig:cos}(a-c), 180$^\circ$ domain walls can be observed in bulk PbTiO$_3$  where Bloch-like polarization components emerge at the domain wall due to their ferroelectric character~\cite{Wojdel-14}.
\begin{figure*}[tb]
     \centering
      \includegraphics[width=15cm]{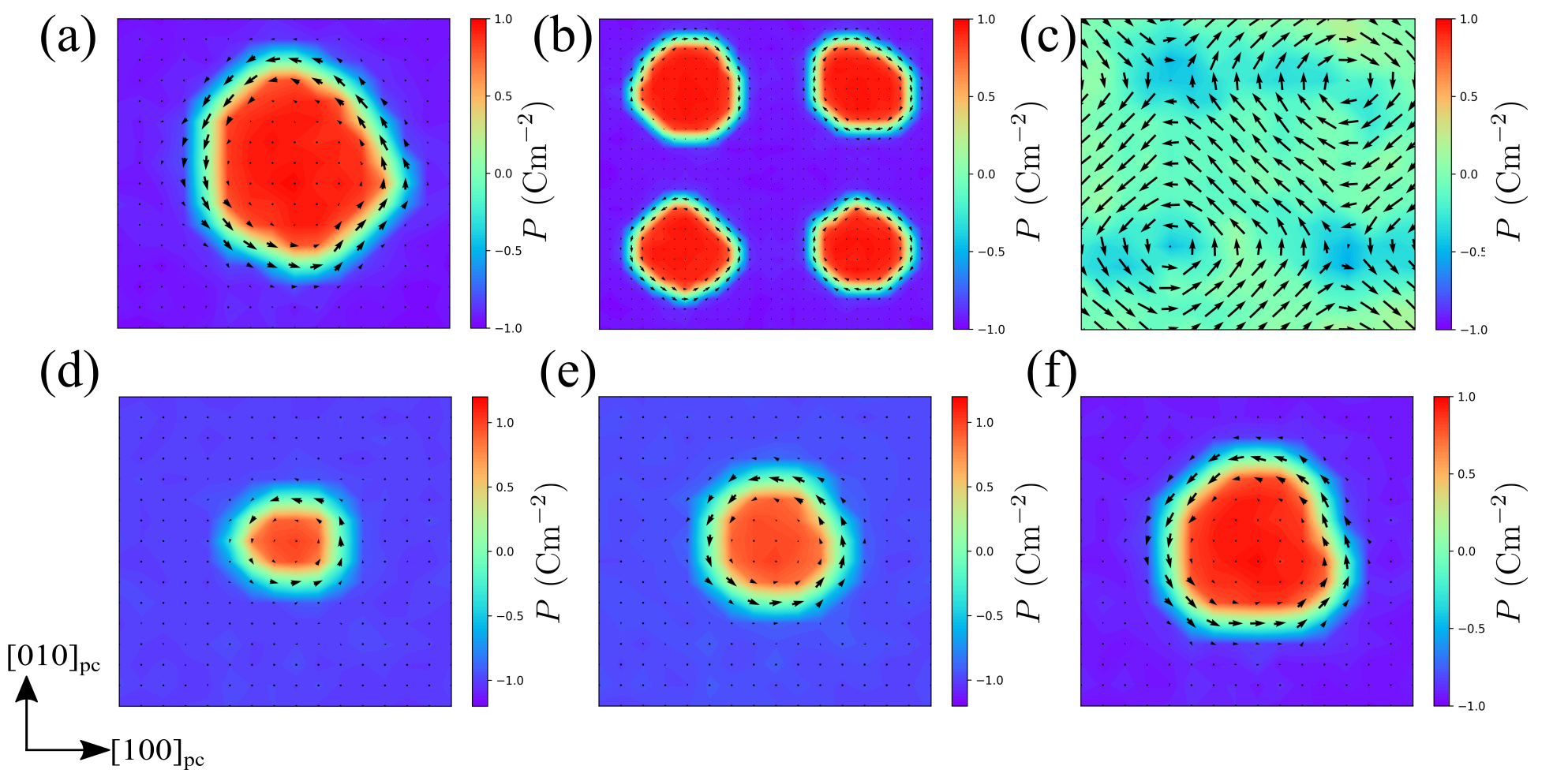}
      \caption{Stabilization of different textures in bulk PbTiO$_3$ by means of different electric fields. (a-b) Skyrmion lattice generated by the superposition of two cosine functions with periodicities of $\lambda_x=\lambda_y=16$ u.c. and $A_z=0.51$ MV/cm in a supercell of size $L=16,~32$ u.c. respectively. (c) Stabilization of a vortex/antivortex lattice in epitaxially strained PbTiO$_3$ $a=b=3.955$ \AA~by the superposition of two cosine functions with periodicities of $\lambda_x=16$, ($\lambda_y=16$) u.c. and $A_y=0.51$, ($A_x=0.51$) MV/cm. (d-f) Individual Skyrmion defects stabilized by the effect of Gaussian pulses as achiavable by an AFM tip with different values of $\sigma=4,~5,~6$ u.c. respectively.}
      \label{fig:bubb} 
\end{figure*}

The polarization configuration obtained from our second-principles calculations does not necessarily mirror the applied electric field, as other relevant energy contributions are inherently at play. As a result, applying electric fields with very narrow periodicities or configurations that promote head-to-head or tail-to-tail domains will not necessarily produce such structures if they lead to energetically unfavorable or unphysical solutions for the system under study.
For example, in bulk PbTiO$_3$, applying a periodicity of $\lambda = 8$ u.c. or smaller does not result in the formation of 180$^\circ$ domain walls with the imposed periodicity. Instead, the dipoles tilt towards the direction perpendicular to the modulation, forming a more uniform pattern that minimizes gradient energies as shown in Fig.~\ref{fig:cos}(d).
Similarly, when ${\bold A}$ and $\bm{\lambda}$ are parallel, the applied electric field naturally favors head-to-head and tail-to-tail domains. For low field magnitudes, these configurations are largely suppressed. As the field strength increases, a slight head-to-head or tail-to-tail configuration may emerge, but the system minimizes the electrostatic energy by lifting the dipoles towards the perpendicular direction to the modulation, thereby reducing the unfavorable bound charges arising from polarization divergence as illustrated in Fig.~\ref{fig:cos}(e). Even at high field magnitudes, the system strives to escape this energetically unfavorable situation.
Likewise, if we apply sufficiently large tensile strain to PbTiO$_3$ ($a=b=3.955$ \AA~ for instance) the regular 180$^\circ$ domains illustrated on Fig.~\ref{fig:cos}(b) further evolve to textures with a non uniaxial polarization pattern. Typically, they adopt a polarization wave arrangement as shown in Fig.~\ref{fig:cos}(f). Interestingly, the PbO domain which previously exhibited a ferroelectric instability along the $z$-direction [see Fig.~\ref{fig:cos}(b)], has now reoriented along the $y$-direction while remaining perpendicular to the external applied field. This solution emerges in order to relax the elastic energy on the system.

\emph{Stabilization of more complex phases.-} 
More complex phases can also be stabilized with other electric fields or by the combination of several waves like the ones used in the previous section. In particular, Skyrmion textures can be stabilized in bulk PbTiO$_3$ [see Fig.~\ref{fig:bubb}(a)] by the superposition of two cosine functions with periodicities along $x$ and $y$ and electric field direction along $z$ similar to the ones used in the previous section. Strictly speaking, this process generates a square lattice of skyrmion bubbles with a periodicity corresponding to that of the initial wave, as illustrated in Fig.~\ref{fig:bubb}(b). By doubling the simulation cell while maintaining the periodicity of the electric field modulation, a square lattice of similar size skyrmions is formed. It is noteworthy that the Bloch component of the domain walls (either clockwise or counterclockwise) varies between individual skyrmions stochastically, demonstrating that they are not identical periodic replicas of one another and granting different chiral nature to each of the defects. 
Upon increasing the tensile strain on the system ($a=b=3.955$ \AA), in-plane lattices of vortices and antivortices as shown in Fig.~\ref{fig:bubb}(c) can be stabilized if two electric fields modulated along $x$ and $y$ and magnitudes along $y$ and $x$ respectively are applied on the system. 
The polarization wave previously stabilized under similar mechanical conditions showing a uniform polarization along the $y$-direction [see Fig.~\ref{fig:cos}(f)] is now unstable due to the effect of the second field modulated along $x$ and with magnitude along $y$. As a consequence, closure textures emerge in the system in order to comply with the imposed periodicity.
\begin{figure*}[tbhp]
     \centering
      \includegraphics[width=17cm]{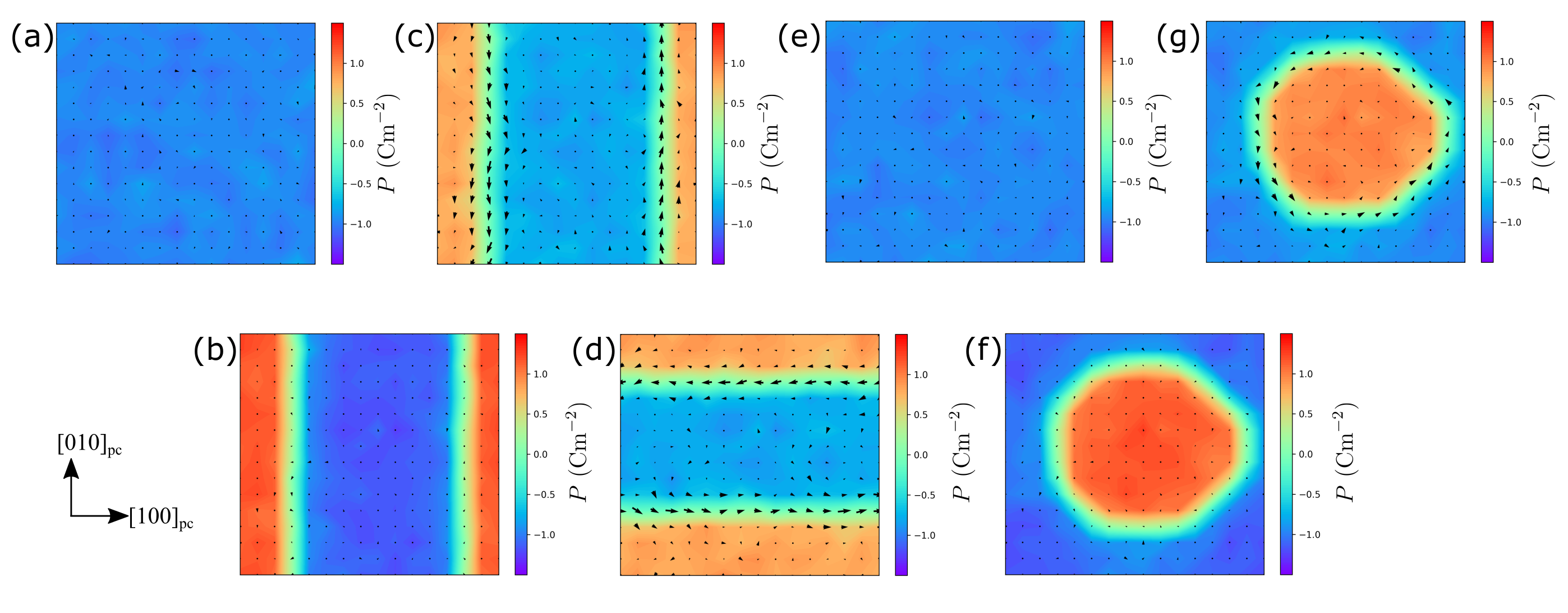}
      \caption{Dynamical evolution of different textures in bulk PbTiO$_3$ stabilized by means of electric fields. (a) Initial phase showing a monodomain configuration. (b) Domain structure under the application of a modulated electric field following Eq.~(\ref{eq:cos}) $\lambda_x=18$ u.c. (c) Relaxed domain strucutre after removing the electric field. (d) Relaxed opposite domain configuration after the application of a modulated field following Eq.~(\ref{eq:cos}) $\lambda_y=18$ u.c.. (e) Monodomain structure obtained under the application of a homogeneous field. (f) Lattice of bubble domains after the application of two modulated fields along $x$ and $y$. (g) Skyrmion lattice after the relaxation of (f) under no external fields.}
      \label{fig:dynamics} 
\end{figure*}
This derives in the evolution of the previously uniform domain wall along the $y$-direction, which now exhibits distinct internal domains one with $P_y > 0$ (up) and the other with $P_y < 0$ (down). Therefore, under the conditions imposed by the modulated electric fields and the tensile strain, a transition from a uniform PbO ferroelectric domain wall~\cite{Wojdel-14} into an Ising line~\cite{Stepkova-15} occurs. Interestingly, these closure textures are stable even if they extend over large distances, allowing their nucleation with electric fields that do not require strong spatial modulation.
The vortex/antivortex singularities are placed at the intersection between the two domain walls and centered on a Pb atom. Interestingly, we observe that an out-of-plane component of the polarization is prone to develop at those points similarly to what happens in regular domain walls~\cite{Wojdel-14} which grants a non-zero topological character to these textures effectively forming a meron/antimeron lattice. 
Similar in-plane vortex/antivortex textures, capable of extending over long distances without requiring highly spatially modulated fields, have already been predicted in BaTiO$_3$ and other rhombohedral ferroelectrics under comparable electric fields~\cite{GomezOrtiz-24}.

Single defects can also be stabilized by the effect of a Gaussian field when applied on a homogeneous background of opposite polarity mimicking the effect that can be achieved by an AFM tip in the laboratory as it was anticipated in Ref.~\cite{Mauro-19} or predicted in Ref.~\cite{Prokhorenko-24}. In a general formulation, such field can be written as
\begin{equation}
    \vec{E}(\mathbf{r})=\mathbf{A}\cdot e^{-(\bm{\mu}-\mathbf{r}-\mathbf{v}t)^2/\sigma^2},
\end{equation}
where $\bm{\mu}$ represents the center of the Gaussian field, $\sigma$ its standard deviation and $\mathbf{v}$ the velocity of the center of mass. 
By selecting different values of $\sigma$, different sizes of bubble domains can be stabilized as it can be observed in Fig.~\ref{fig:bubb}(d-f). However, as discussed in the previous section, if lower values of $\sigma$ are chosen, the applied field will be unable to nucleate any domain unless unrealistically large field amplitudes are employed. In such cases, the resulting domains, will be unstable and collapse upon a monodomain phase after removing the bias.

\emph{Reversible and deterministic switching.-} 
One of the strengths of stabilizing different polar textures via applied electric fields, compared to conventional methods such as strain engineering~\cite{Dai-23} or varying growth conditions~\cite{Hong-17}, is the ability to actively manipulate the polar state of the material, enabling real-time control
This section presents concrete examples where domains along different directions or skyrmion lattices can be stabilized and dynamically switched between each other by varying the applied electric field to which the system is subjected.
To illustrate this, Fig.~\ref{fig:dynamics} collects the evolution of the system under finite temperatures of $T=70$ K when subjected to different electric fields. 

Initially [Fig.~\ref{fig:dynamics}(a)], we start with a monodomain structure obtained after poling the system with an homogeneous electric field. Then, we nucleate a domain structure periodically repeated along the $x$-direction after the application of a field satisfying Eq.~(\ref{eq:cos}) with $\lambda_x=18$ u.c. and $A_z=5.14$ MV/cm [Fig.~\ref{fig:dynamics}(b)] that relaxes upon removal of the electric field [Fig.~\ref{fig:dynamics}(c)]. Such domain can be switched towards the $y$-direction after the application of a softer field showing periodicity of $\lambda_y=18$ u.c. and $A_z=1.54$ MV/cm. Notably, these field-induced phases are long-lived metastable states and remain stable even after the field is removed [Fig.~\ref{fig:dynamics}(d)].
\begin{figure*}[tbhp]
     \centering
      \includegraphics[width=15cm]{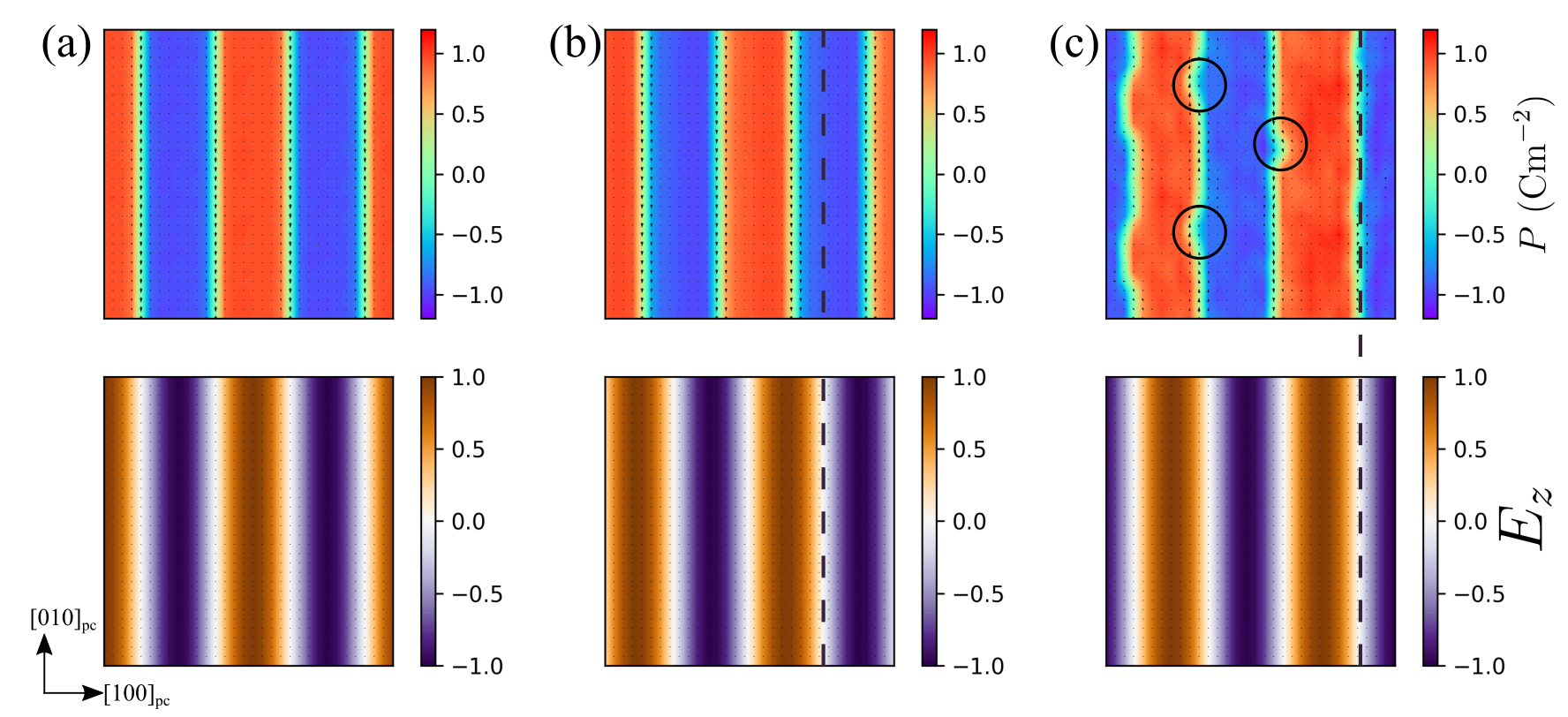}
      \caption{Domain wall motion in bulk PbTiO$_3$. Domain structure (upper row) and instantaneous electric field (bottom row) at (a) $t=0$ s. (b) $t=4.8$ ns and (c) $t=9.3$ ns. Arrows in the dipole patterns represent the in-plane components of the polarization. The color bar in the electric field represents its normalized value along the $z$ direction. The dashed lines in panels (b) and (c) represents the location of the node of the electric field wave and the circles in panel (c) highlight domain nucleation points.}
      \label{fig:move_DW} 
\end{figure*}
This domain structures can be erased (by applying homogeneous electric fields $A_z=1.54$ Mv/cm) or rewritten back and forth at will following the previously described procedure. Moreover, we can transit between periodically repeated domain structures and lattices of Skyrmion if instead of a single field modulated along $x$ or $y$ we applied both fields at the same time as shown in the previous section $\lambda_{1,x}=\lambda_{2,y}=18$ u.c. of magnitude $A_{z}=5.14$ MV/cm [Fig.~\ref{fig:dynamics}(f)]. After removing the field, the periodic lattice of bubble nanodomains further relaxes to a lattice of skyrmion textures demonstrating again the stability of the nucleated phases and their deterministic control, similar to that achieved via APEX~\cite{Bastogne-24}.    

\emph{Dynamics of ferroelectric domains.-} 
After stabilizing domain structures and demonstrating deterministic switching between different configurations, we now investigate how domain walls can be displaced using electric fields. This section focuses on their mobility, examining how domain walls respond to changes in the electric field and how fast they can rearrange.
In order to account with the time modulation of the electric field an extra term needs to be added to the functional form presented in Eq.~(\ref{eq:cos}) that now reads
\begin{equation}
        {\bold E}({\bold r})=\mathbf{A}\cdot \cos(\frac{2\pi x}{\lambda_x}+\frac{2\pi y}{\lambda_y}+\frac{2\pi z}{\lambda_z}+\frac{2\pi t}{T}),
    \label{eq:cos_time}
\end{equation}
where $T$ is the period of the time-modulation. The expected velocity of the electric field wave, corresponding to the propagation speed of its phase, is then given by $v=\left(1/\lambda_x^2+1/\lambda_y^2+1/\lambda_z^2\right)^{-1/2}/T$, which depends on the spatial and temporal periodicities and that in the most usual case of a single spatial modulation reduces to $v=\lambda/T$.

The motion of domain walls is critical to many applications involving ferroelectric materials and has been deeply studied in the context of homogeneous electric fields under the nucleation and growth framework~\cite{Merz-54,Miller-60,Shin-07,Paruch-02,Paruch-13}. However, their behavior under inhomogeneous electric fields, as well as the dynamics of topological textures, remains comparatively underexplored, with some notable exceptions in the literature~\cite{Prokhorenko-24,Jorge-24}.
In Fig.~\ref{fig:move_DW}, we show the molecular dynamics results for the domain wall motion under an external electric field of magnitude $2$ MV/cm, periodicity $\lambda_x=16$ u.c. and velocity $v=300$ m/s. 

A primary and evident difference compared to the nucleation and growth mechanism under homogeneous fields is that, in our case, the electric field varies dynamically with time. This temporal variation imposes a constraint on the domain walls, requiring them to adapt to the periodic changes in the electric field. However, as Fig.~\ref{fig:move_DW}(b) reveal, a noticeable lag in their response to the field's oscillations can be observed (see dashed line), proving a delay in the reaction time of the domains to the external driving force that amounts to 4 ps at $T=5$ K for the 16 u.c. domain periodicity. Interestingly, this lag is only present at the onset of the dynamics; once the domains begin to move, the correlation between the domain wall position and the node of the electric field wave becomes one-to-one [see dashed line in Fig.~\ref{fig:move_DW}(c)]. Moreover, this lag shows a strong temperature dependence, when the system is heated to 100 K, the lag is reduced by half, and at 200 K, it already becomes negligible. Interestingly, when the electric field is removed, the domains stop moving immediately, showing no sign of residual momentum, although this observation can be slightly influenced as a consequence of the domain propagation mechanism.
Our MD simulations demonstrate that domain displacement follows the diffuse boundary model as proposed in Ref.~\cite{Shin-07}. As shown by the solid circles in Fig.~\ref{fig:move_DW}, certain regions of the domain nucleate first, and the domain wall subsequently follows this initial nucleation, rather than propagating uniformly across the domain. Therefore, if the nucleus of the new domain exceeds a critical size, it can lead to the formation of an additional domain layer even after the electric field is removed. However, this phenomenon is more related to the relaxation of the domain structure than to an inertial response.
While the absence of inertia upon field removal is consistent with the findings of Ref.~\cite{Liu-13}, our results complement them by demonstrating a clear inertial response at the initiation of domain wall motion. This asymmetry in the inertial behavior of domains comes from the intrinsic energy barrier for nucleation.

Besides, there are scenarios where domains remain immobile due to the high activation energy or inertia associated with their motion irrespective of the wave velocity. Our observations indicate, for instance, that domain mobility is strongly influenced by their size: domains with double the periodicity ($32$ u.c.) remain static  when the electric field is adjusted to match their modulation but retains the same magnitude that renders domains with a periodicity of $16$ u.c. already mobile. 
However, the activation energy barrier can be mitigated by either increasing the temperature or enhancing the magnitude of the electric field applied to the domains. Specifically, when the periodicity of the domains and the corresponding electric fields is doubled, the field's magnitude should also be doubled in order to exert sufficient force on the domains, enabling their movement. This relationship between activation energy, temperature, and electric field is consistent with the predictions of the creep law~\cite{Paruch-02,Paruch-13}. These adjustments enable domains to transition from an immobile to a mobile regime, allowing even larger domains to overcome their inherent inertia. 

When the propagation speed of the electric field increases, the domain structure is unable to keep up with the rapid oscillations and eventually regions of negative electric field align with positive polarization regions. This misalignment induces a chaotic evolution in the system, leading to the fragmentation of the domain structure and marking the maximum velocity at which domains can move. In our case, with $\lambda_x = 16$ u.c. and a field amplitude of $2$ MV/cm, this limiting velocity was approximately $3000$ m/s, significantly higher than the $900$ m/s recently reported for magnetic skyrmions~\cite{Pham-24}. If the wave velocity of the electric field is increased further, there is a threshold beyond which the domains become effectively immobile, no longer responding to the applied electric field.
Interestingly, the limiting velocity at which domains lose connectivity and destabilize remains unaffected by changes in the electric field magnitude or temperature. The domain structure consistently destabilizes at the same electric field velocity, marking an intrinsic upper bound on domain wall motion that is probably related with the material and the stability of the multidomain structure.

Similar characteristics, lag in response and asymmetric inertial responses, the existence of a limiting velocity (which is indeed identical to the one previously reported), and the diffuse boundary nucleation, are also observed in the dynamics of bubble domains. In such a case, the motion of the bubble domains is governed by the elongation and contraction of their structure, resembling the deformation of a water droplet or the dynamic movement of an amoeba.
Remarkably, the mobility of these bubble domains is isotropic, enabling them to move equivalently in any direction in response to the position of a Gaussian field, without exhibiting any preference for specific crystallographic directions. Furthermore, they can adapt to other waveforms, such as cosine functions, maintaining their structural integrity while being displaced by these periodic perturbations under appropriate electric field magnitudes. 
\section{Conclusions}
In this work, we demonstrate the deterministic nucleation and precise control of various polar textures in the prototypical ferroelectric PbTiO$_3$ using spatially modulated electric fields. 
In contrast to regular methods for the synthesis of topological textures that do not allow the modification of the polar arrangement once it is stabilized~\cite{Hong-17,Dai-23}, our methodology allows to dynamically manipulate polar textures in situ, holding promise for the development of next-generation nanoelectronic devices, where precise control over different ferroelectric phases is crucial.

Furthermore, we have investigated the motion of multidomain structures under inhomogeneous electric fields and observed a noticeable lag between the oscillations of the electric field and the response of the polar texture. This lag, which arises from the inertia of the polarization domains, is particularly evident at the onset of the dynamics and shows a strong temperature dependence but becomes negligible once the domains achieve steady motion. Interestingly, while the domains exhibit an inertial response at the onset of motion, it is not present upon the cessation of the field, with the domains coming to rest immediately after field removal and showing no residual momentum.

As the velocity of the electric field increases, the synchronization between the electric field and the domain structure is lost, ultimately leading to the fragmentation of the domain configuration. Notably, this limiting velocity is remarkably robust against variations in the electric field magnitude and temperature. However, it presumably  depends on the material properties or the intrinsic stability of the domains. For instance, in ferroelectric/dielectric superlattices, where domains are inherently formed due to electrostatic conditions, the limiting velocity will likely be larger.
Additionally, our simulations highlight the significant role of domain size in their mobility under inhomogeneous fields. Larger domains exhibit a higher activation energy barrier, rendering them immobile under conditions where smaller domains are already mobile.

These findings provide valuable insights into the intricate dynamics of domain walls and topological textures under inhomogeneous electric fields, shedding light on the fundamental mechanisms governing their motion.
\acknowledgments
F.G.O. acknowledges financial support from MSCA-PF 101148906 funded by the European Union and the Fonds de la Recherche Scientifique (FNRS) through the grant FNRS-CR 1.B.227.25F and the Consortium des Équipements de Calcul Intensif (CÉCI), funded by the F.R.S.-FNRS under Grant No. 2.5020.11 and the Tier-1 Lucia supercomputer of the Walloon Region, infrastructure funded by the Walloon Region under the grant agreement No. 1910247. F.G.-O. and Ph. G. also acknowledge support by the European Union’s Horizon 2020 research and innovation program under Grant Agreement No. 964931 (TSAR).  Ph. G. and X.H. also acknowledges support from the Fonds de la Recherche Scientifique (FNRS) through the PDR project PROMOSPAN (Grant No. T.0107.20).
%
\end{document}